\documentclass{article}

\usepackage{microtype}
\usepackage{booktabs} 

\usepackage{hyperref}

\usepackage{amssymb}
\usepackage{amsmath}
\usepackage{algorithm}
\usepackage{algorithmicx}
\usepackage{algpseudocode}
\usepackage{enumitem}
\usepackage{pifont}
\usepackage[normalem]{ulem}
\usepackage{xcolor}
\usepackage{lipsum}
\usepackage[htt]{hyphenat}
\usepackage{seqsplit}

\usepackage{graphicx}
\usepackage{caption}

\usepackage{url}
\usepackage{listings}
\usepackage{subcaption}

\definecolor{graybg}{RGB}{248,249,250} 
\definecolor{bordergray}{RGB}{206,212,218}
\definecolor{darkgreen}{RGB}{40,167,69}

\lstdefinestyle{elegantbash}{
    language=bash,
    backgroundcolor=\color{graybg},
    basicstyle=\ttfamily\small,
    keywordstyle=\color{blue}\bfseries,
    commentstyle=\color{darkgreen},
    stringstyle=\color{orange},
    showstringspaces=false,
    breaklines=true,
    breakatwhitespace=false,
    breakindent=20pt,
    frame=single,               
    rulecolor=\color{bordergray},  
    frameround=tttt,           
    framesep=3pt,             
    xleftmargin=0.5cm,
    xrightmargin=0.5cm,
    aboveskip=1.5em,
    belowskip=1.5em
}

\usepackage[accepted]{mlsys2025}

\NewDocumentCommand{\var}{O{s} m O{}}{%
  \ensuremath{#1_{#2}^{#3}}
}
\usepackage{siunitx}
\usepackage{colortbl}
\usepackage{xspace}

\newcommand{\commentout}[1]{}

\definecolor{light-gray}{gray}{0.80}

\newcommand\aref{Algorithm~\ref}

\newcommand\fref{Fig.~\ref}
\newcommand\tref{Table~\ref}
\newcommand\sref{\S~\ref}

\usepackage{amsthm}

\usepackage{amsthm}

\newtheorem{mytheorem}{Theorem}
\newtheorem{assumption}[mytheorem]{Assumption}
\newtheorem{lemma}[mytheorem]{Lemma}
\newtheorem{remk}[mytheorem]{Remark}

\newcommand{\hide}[1]{}

\renewcommand{\emph}[1]{\textit{#1}}

\newcommand{\name}{SuperInfer\xspace}

\newcommand{\scheduler}{RotaSched\xspace}

\newcommand{\swapper}{DuplexKV\xspace}

\usepackage[dvipsnames]{xcolor}

\definecolor{darkgreen}{RGB}{0,165,0}

\newcommand{\revised}[1]{#1}

\mlsystitlerunning{\name: SLO-Aware Rotary Scheduling and Memory Management for LLM Inference on Superchips}

\begin{document}

\twocolumn[
\mlsystitle{\name: SLO-Aware Rotary Scheduling and Memory Management for LLM Inference on Superchips}



\mlsyssetsymbol{equal}{*}

\begin{mlsysauthorlist}
\mlsysauthor{Jiahuan Yu}{uiuc}
\mlsysauthor{Mingtao Hu}{uiuc}
\mlsysauthor{Zichao Lin}{uiuc}
\mlsysauthor{Minjia Zhang}{uiuc}
\end{mlsysauthorlist}

\mlsysaffiliation{uiuc}{Siebel School of Computing and Data Science, University of Illinois Urbana-Champaign, Champaign, USA}

\mlsyscorrespondingauthor{Jiahuan Yu}{jiahuan2@illinois.edu}
\mlsyscorrespondingauthor{Mingtao Hu}{mingtao4@illinois.edu}
\mlsyscorrespondingauthor{Zichao Lin}{zichaol3@illinois.edu}
\mlsyscorrespondingauthor{Minjia Zhang}{minjiaz@illinois.edu}

\mlsyskeywords{Machine Learning, MLSys}

\vskip 0.3in

\begin{abstract}
Large Language Model (LLM) serving faces a fundamental tension between stringent latency Service Level Objectives (SLOs) and limited GPU memory capacity.
When high request rates exhaust the KV cache budget, existing LLM inference systems often suffer severe head-of-line (HOL) blocking.
While prior work explored PCIe-based offloading, these approaches cannot sustain responsiveness under high request rates, often failing to meet tight Time-To-First-Token (TTFT) and Time-Between-Tokens (TBT) SLOs. 
We present \name, a high-performance LLM inference system designed for emerging Superchips (e.g., NVIDIA GH200) with tightly coupled GPU-CPU architecture via NVLink-C2C.
\name introduces \scheduler, the first proactive, SLO-aware rotary scheduler that rotates requests to maintain responsiveness on Superchips, and \swapper, an optimized rotation engine that enables full-duplex transfer over NVLink-C2C.
Evaluations on GH200 using various models and datasets show that \name improves TTFT SLO attainment rates by up to 74.7\% while maintaining comparable TBT and throughput compared to state-of-the-art systems, demonstrating that SLO-aware scheduling and memory co-design unlocks the full potential of Superchips for responsive LLM serving. 
Code is available in \url{https://github.com/Supercomputing-System-AI-Lab/SuperInfer}.
\end{abstract}

]



\printAffiliationsAndNotice{}  



\section{Introduction}
\label{sec:intro}

Large Language Models (LLMs)~\cite{achiam2023gpt,touvron2023llama,liu2024deepseek,bai2023qwen} increasingly power latency-critical applications.
These services demand strict latency \emph{Service Level Objectives} (SLOs), particularly \emph{Time-To-First-Token} (TTFT) and \emph{Time-Between-Tokens} (TBT), to ensure responsive user interactions.
Consequently, building high-performance LLM serving systems capable of meeting SLOs while maintaining high throughput has become a critical research challenge.

The key difficulty lies in memory-intensive inference.
Each request maintains an ever-growing key-value (KV) cache during autoregressive generation~\cite{radford2019language}.
Under high request rates, this cumulative KV cache quickly exhausts GPU memory, leading to severe \emph{head-of-line} (HOL) blocking and SLO violations.
While prior SLO-aware schedulers~\cite{agrawal2024taming,fu2024efficient,hongsola} reduce queuing delays via priority reordering, they are ultimately constrained by limited on-device memory, failing to serve requests whose KV cache no longer fits.

To overcome this bottleneck, a line of work explores \emph{offloading}, which swaps KV cache or model parameters to CPU memory~\cite{sheng2023flexgen,jiang2024neo,hu2024memserve}.
While expanding effective memory capacity, existing techniques suffer two critical limitations.
First, most of them are SLO-unaware, reacting to memory pressure rather than latency urgency, often causing severe TBT SLO violations for requests stuck in swapped queues (\sref{subsec:obs-why-need-scheduler}).
Second, they are typically designed for PCIe architectures, whose limited bandwidth ($\sim$32-64GB/s) makes swapping too slow to alleviate HOL blocking under high load (\sref{subsec:obs-pcie-vs-gh200}).

The emergence of GPU-CPU tightly-coupled Superchips, such as the NVIDIA GH200~\cite{gh200}, fundamentally shifts this landscape.
GH200 integrates a Hopper GPU and a Grace CPU via NVLink-C2C, a cache-coherence interconnect with up to 900GB/s bandwidth, an order of magnitude higher than PCIe.
While initial studies explore GH200 for LLM serving~\cite{xu2024pie} and training~\cite{lian2025superoffload}, its full potential remains untapped.
Specifically, directly porting existing PCIe-based offloading mechanisms to GH200 yields unexpectedly poor results due to $<$5\% utilization of C2C bandwidth (\sref{subsec:obs-gh200-challenges}).
This counterintuitive result suggests the bottleneck lies in the software stack instead of hardware.
What prevents today's serving stacks from exploiting the full potential of Superchips, and what design principles are needed to close this gap?

Our in-depth study of GH200's memory hierarchy and offloading behavior reveals fundamental software-hardware mismatches preventing current serving stacks from leveraging its full potential (\sref{subsubsec:why-under-utilization}).
Guided by our observations, we introduce \name, a high-performance, SLO-aware LLM serving system optimized for Superchips.
\name employs two co-designed techniques for responsive, full-duplex offloading: 
(1) \textbf{\scheduler} (\sref{subsec:design-scheduler}), an OS-inspired rotary scheduler that, instead of using passive preemption for out-of-memory error prevention, introduces a novel transient \emph{rotary} state and \emph{active rotation} to proactively rotate requests between HBM and DRAM based on their SLO progress.
(2) \textbf{\swapper} (\sref{subsec:design-swapper}), a high-performance KV cache rotation engine co-designed to support efficient request rotations over NVLink-C2C.
It maximizes C2C bandwidth utilization by enabling full-duplex data-race-free transfers, merging fragmented KV cache segments into large efficient batches, and overlapping transfers with model execution.

We evaluate \name across various models and workloads on GH200.
Under high request rates, \name substantially improves end-to-end SLO attainment, achieving up to 74.7\% higher TTFT success rates than state-of-the-art systems, while maintaining comparable throughput and TBT.
At low request rates, where memory is sufficient, \name matches baseline performance, confirming its benefits stem from efficient, SLO-aware memory management on Superchips.
These improvements demonstrate that Superchips, paired with co-designed software, can effectively mitigate HOL blocking and enable responsive LLM serving.

In summary, this paper makes the following contributions:
\textbf{(1)} We identify the limitations of existing LLM serving systems with PCIe-based offloading on Superchips, and analyze why they fail to utilize NVLink-C2C bandwidth and achieve SLO responsiveness.
\textbf{(2)} We present \name, the first SLO-aware LLM serving system co-designed for Superchips, which consists of (i) \scheduler, an OS-inspired scheduler using active rotation guided by Virtual Lag Time (VLT), and (ii) \swapper, a full-duplex KV cache rotation engine that eliminates C2C under-utilization.
\textbf{(3)} We demonstrate that \name achieves up to 74.7\% higher TTFT SLO attainment rate under high request rates while maintaining comparable throughput and TBT latency.




\section{Background and Related Work}
\label{sec:bg}

\subsection{LLM Inference and Offloading over PCIe GPUs}
\label{subsec:bg-llm-inference-and-offload}

During LLM autoregressive generation, each request's KV cache grows linearly with sequence length~\cite{radford2019language}, making GPU memory the dominant bottleneck: even high-end GPUs can hold only a limited number of concurrent requests.
Researchers have explored sophisticated mechanisms to address this challenge.

Early systems like DeepSpeed-Inference~\cite{aminabadi2022deepspeed} and FlexGen~\cite{sheng2023flexgen} treat host memory as an extension of GPU memory to fit models beyond on-device capacity.
While enabling single-GPU inference for large models, their static offloading leads to high data transfer latency during dynamic serving workloads.
Later, PagedAttention~\cite{kwon2023efficient} proposes paging-based memory managers that organize KV cache into small, fixed-size blocks, significantly reducing memory fragmentation and improving memory utilization.
This design has become the de facto technique in recent LLM serving frameworks~\cite{kwon2023efficient,zheng2024sglang}.
However, paging scatters KV cache across non-contiguous small regions, making GPU-CPU transfer expensive.
Subsequent systems attempt to overlap the transfer overheads with computation~\cite{jiang2024kvpr,luo2025headinfer,cao2025moe,yu2024twinpilots,kim2025lia,hu2024memserve,qin2025mooncake}.
FastDecode~\cite{he2024fastdecode},
NEO~\cite{jiang2024neo},
HeteGen~\cite{zhao2024hetegen},
CachedAttention~\cite{gao2024cost},
FlashGen~\cite{jeong2025accelerating},
and NanoFlow~\cite{zhu2025nanoflow} propose offloading KV cache or part of the computation from GPU to CPU or SSDs, effectively increasing the batch size and therefore inference throughput. 
Although effective, these designs remain PCIe-bound, and do not exploit the high-bandwidth, full-duplex C2C links in emerging Superchips.

To overcome the PCIe bottleneck, Aqua~\cite{vijaya2025aqua} offloads KV cache to other GPUs via inter-GPU NVLink, but it relies on limited spare peer GPU memory rather than large CPU memory on Superchips.
Alternatively, some systems reduce KV cache volume itself instead of transfer pipeline optimizations~\cite{chen2025impress,song2024powerinfer}.
For example, InfiniGen~\cite{lee2024infinigen} prunes KV entries via dynamic importance estimation and transfers only a subset of KVs for each request.
CacheGen~\cite{liu2024cachegen} compresses KV cache via layer-wise quantization and arithmetic coding.
These methods reduce offloading overhead but are lossy, making them hard to generalize to unseen tasks and requests.

\subsection{Offloading over Emerging Tightly Coupled GPU–CPU Architectures}
\label{subsec:bg-offloading-gh200}

Recent Superchip architectures (e.g., NVIDIA GH200, AMD MI300A) tightly integrate GPU and CPU via high-performance interconnects, offering an order of magnitude higher bandwidth than traditional PCIe.
For example, GH200's NVLink-C2C provides 900GB/s bidirectional bandwidth~\cite{gh200}.
Several recent studies have benchmarked~\cite{fusco2024understanding,schieffer2024harnessing} or explored programming Superchips~\cite{vellaisamy2025characterizing}.
SuperOffload~\cite{lian2025superoffload} further examines Hopper, Grace, C2C joint optimizations, demonstrating potential for large-scale LLM training.
However, few prior systems explicitly target LLM inference on Superchips. 
The most relevant work is Pie~\cite{xu2024pie}, which enables GPU-CPU KV cache spilling on GH200 and adapts memory allocation on-the-fly.
However, Pie is not optimized for tight latency SLOs under high load, nor does it apply SLO-aware scheduling.
In contrast, \name introduces an active rotary scheduler and full-duplex transfer engine, jointly addressing SLO-awareness and C2C link utilization.

\subsection{Scheduling for Latency-Aware Inference}
\label{subsec:bg-latency-aware-scheduler}

While widely used techniques like continuous batching~\cite{yu2022orca} improve throughput, they lack explicit latency control.
Several works target fairness among requests or clients~\cite{sheng2024fairness,wei2025equinox}, which complement but do not directly optimize for latency SLOs.

There has been a growing number of works explicitly managing request latency.
Sarathi-Serve~\cite{agrawal2024taming} introduces chunked prefill and a dynamic batching policy to improve SLO attainment under dynamic workloads, but it assumes GPU-residency of all requests.
\citet{ao2025optimizing} formulate LLM serving as an online scheduling problem and proposes WAIT/Nested-WAIT algorithms to balance throughput and latency under memory constraints. 
While effective within single-GPU memory, these works do not address latency management of LLM serving for heterogeneous GPU-CPU hierarchies or offloading.
A larger set of systems mitigates latency via queue reordering and request prioritization.
For example, SSJF~\cite{qiu2024efficient} and LTR~\cite{fu2024efficient} use length prediction to approximate Shortest-Job-First (SJF) or Shortest-Remaining-Time-First (SRTF) scheduling, reducing HOL blocking and improving responsiveness of short requests.
Others treat SLO attainment as a search or scheduling optimization problem, such as SOLA~\cite{hongsola}, FastServe~\cite{wu2023fast}, SHEPHERD~\cite{zhang2023shepherd}, or target multi-SLO serving for mixed request types~\cite{zhang2025tempo,huang2025slo,borui2025efficient,li2025adaserve}.
Although effective, they are fundamentally limited by GPU memory capacity.

Alternatively, several works aim to reduce SLO violations by saving GPU memory or increasing utilization~\cite{chen2025multiplexing,patke2024queue}.
LightLLM~\cite{gong2025past} estimates future KV cache usage to avoid harmful preemption, which wastes GPU memory, but it is also limited by GPU memory capacity.
TokenFlow~\cite{chen2025tokenflow} and Select-N~\cite{ma2025memory} explore SLO-aware scheduling with offloading to balance host memory usage and latency, which are PCIe-based and thus limited by its bandwidth.

\section{Observations}
\label{sec:obs}

This section presents the motivational studies.
We first identify the SLO-aware offloading as a challenge (\sref{subsec:obs-why-need-scheduler}), then examine the limitations of PCIe for offloading (\sref{subsec:obs-pcie-vs-gh200}), finally analyze the unexpected NVLink-C2C bandwidth under-utilization of existing offloading techniques (\sref{subsec:obs-gh200-challenges}).


\subsection{SLO-Aware Offloading as a Challenge}
\label{subsec:obs-why-need-scheduler}

\begin{figure}[!t]
    \centering
    \captionsetup{skip=1pt}
    \includegraphics[width=\columnwidth]{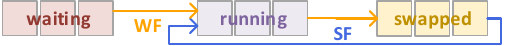}
    \\[0.7em]
    \includegraphics[width=\columnwidth]{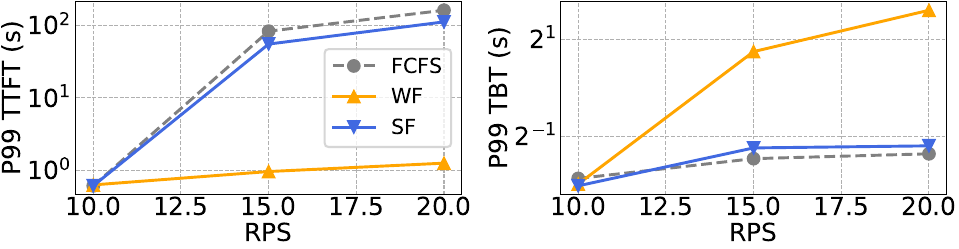}
    \caption{Two static offloading policies: Waiting-First (WF) and Swapped-First (SF), and comparison of their P99 TTFT and TBT to FCFS under varying request rates (Qwen2.5-32B, ShareGPT).}
    \label{fig:obs-schedule-problem-latency}
\end{figure}

While sophisticated scheduling techniques exist, they cannot fundamentally overcome the GPU memory pressure imposed by hardware constraints (more details in Appendix~\ref{app:sjf}).
A natural idea to mitigate memory pressure is offloading, which swaps inactive requests' KV cache to CPU memory.
However, this introduces a challenge for SLO-aware serving (\fref{fig:obs-schedule-problem-latency}): prioritize waiting requests (\textit{Waiting-First}, WF) by preempting running ones, or prioritize resuming swapped requests (\textit{Swapped-First}, SF)?
These static, SLO-unaware policies, common in production frameworks~\cite{zheng2024sglang,kwon2023efficient}, fail to strike a balance.
As \fref{fig:obs-schedule-problem-latency} shows, compared to \textit{First-Come-First-Serve} (FCFS), WF favors new arrivals, reducing TTFT but significantly increasing TBT, as running requests are paused for a long time during generation.
Conversely, SF protects TBT but degrades to FCFS-like performance, failing to fully exploit the potential of offloading, as swapped requests are prioritized to resume immediately, thus swap space is underutilized.
Consequently, neither policy effectively balances TTFT and TBT. 




\textbf{\uline{Insight \#1}}: Offloading shows promise for alleviating memory pressure, but making it SLO-aware requires determining \emph{which} requests to swap or run.
Static policies inevitably bias toward one SLO at the expense of the other.
An effective design must mitigate this issue via proactive scheduling.

\subsection{Responsiveness Bottleneck Caused by Low Swap Bandwidth over PCIe}
\label{subsec:obs-pcie-vs-gh200}

Even with ideal scheduling, SLO-aware offloading succeeds only if requests can be swapped fast enough between GPU and CPU.
In practice, offloading responsiveness is constrained by the effective \emph{swap bandwidth}, i.e., the amount of KV cache transferred per unit time, which, on PCIe-based GPUs, is fundamentally limited by the bandwidth of PCIe.

\begin{figure}[!t]
    \centering
        \begin{minipage}[b]{0.48\columnwidth}
        \centering
        \captionsetup{skip=1pt}
        \includegraphics[width=\linewidth]{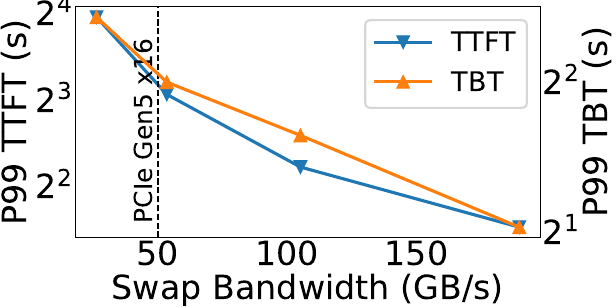}
        \caption{P99 TTFT and TBT latency vs. swap bandwidth for vLLM with offloading. (Qwen2.5-32B, ShareGPT, RPS=20).}
        \label{fig:bg-pcie-swap-budget}

    \end{minipage}
    \hfill
    \begin{minipage}[b]{0.48\columnwidth}
        \centering
        \includegraphics[width=\linewidth]{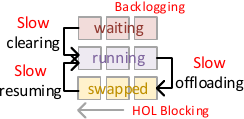}
        \caption{Limitations of low swap bandwidth for offloading: slow clearing of backlogged waiting requests and new HOL blocking in the swapped queue.}
        \label{fig:bg-pcie-new-hol}
    \end{minipage}
\end{figure}

As \fref{fig:bg-pcie-swap-budget} shows, increasing swap bandwidth beyond the PCIe Gen5x16 uni-directional limit significantly reduces both TTFT and TBT.
Tail TTFT is primarily caused by waiting queue delay, and tail TBT primarily results from the duration that requests are in the swapped state.
Thus, PCIe's low swap bandwidth creates two major obstacles to reducing tail latencies (\fref{fig:bg-pcie-new-hol}).
The first, it leads to \emph{request backlogging}, i.e., the accumulation of new arriving requests in the waiting queue due to GPU memory exhausting at high request rates.
Low swap bandwidth makes clearing this backlog through offloading sluggish; paused requests remain in GPU memory longer due to slow offloading, while new arrivals continue to queue up, increasing TTFT SLO violations. 
Second, when a large number of requests are swapped, low swap bandwidth stalls request resumption  due to slow KV cache transfer, leading to new HOL blocking in the swapped queue and limited responsiveness to TBT SLO violations.
Consequently, PCIe fundamentally limits ability of offloading-based LLM serving systems to respond to dynamic SLO violations when system load is high.

\vspace{-0pt} 

\textbf{\uline{Insight \#2}}:
Offloading-based SLO-aware LLM inference is fundamentally limited by low bandwidth of PCIe.
It constrains the swap bandwidth and responsiveness, leading to slow clearing of backlogged waiting requests and new HOL blocking in the swapped queue.

\subsection{Superchip Offers Opportunity but Not for Free}
\label{subsec:obs-gh200-challenges}

In principle, the GH200's high NVLink-C2C bandwidth could eliminate the swap bandwidth bottleneck described above, enabling near-instantaneous request preemption and resumption.
However, our profiling of widely-used LLM serving systems shows severe bandwidth under-utilization.

\begin{figure}[!t]
    \centering    
    \begin{minipage}[b]{0.48\columnwidth}
        \centering
        \includegraphics[width=\columnwidth]{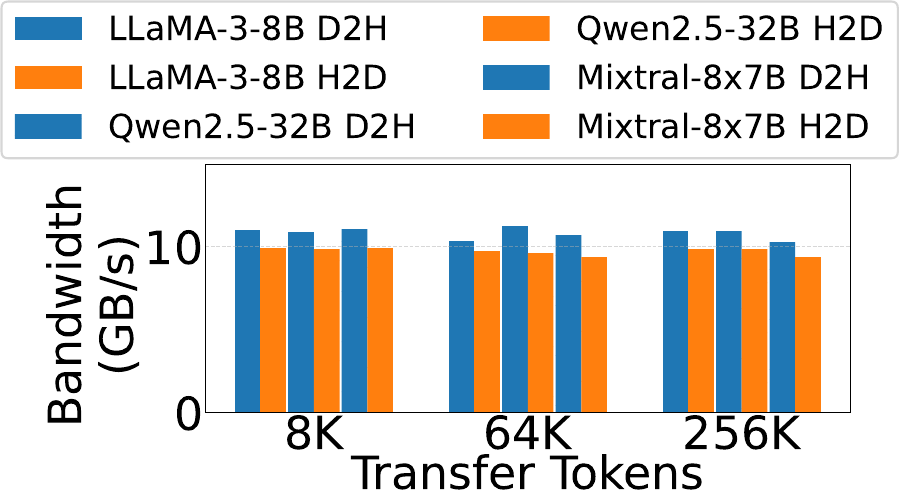}
        \caption{Measured bidirectional CPU-GPU KV cache transfer bandwidth of vLLM for three different models.}
        \label{fig:obs-bw}
    \end{minipage}
    \hfill
    \begin{minipage}[b]{0.48\columnwidth}
        \centering
        \includegraphics[width=\linewidth]{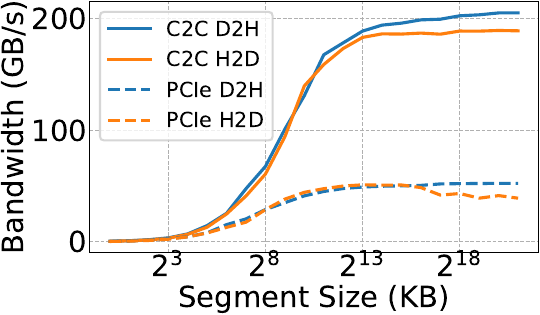}
        \caption{NVLink-C2C (GH200) vs. PCIe Gen5x16 (H200) full-duplex bandwidth by segment size.}
        \label{fig:bg-bw}
    \end{minipage}
\end{figure}

As \fref{fig:obs-bw} shows, vLLM's offloading engine achieves only $\sim$10GB/s of effective bandwidth on GH200 ($<$5\% of the theoretical peak) across three models, regardless of the total transfer token number.
To understand this gap, we perform a fine-grained full-duplex transfer bandwidth characterization across different segment sizes for GH200 (see Appendix \ref{app:bw-measure} for method).
As \fref{fig:bg-bw} shows, C2C bandwidth saturates at $\sim$200+200GB/s.
Further investigation shows this limit stems not from the C2C itself, but from the Grace DRAM subsystem.
Each NUMA node in Grace provides 384GB/s of DRAM bandwidth~\cite{gh200}.
Therefore, memory-intensive transfers can saturate DRAM bandwidth before reaching the C2C link's 450+450 GB/s theoretical peak.
Moreover, Grace DRAM is half-duplex, which explains why concurrent device-to-host (D2H) and host-to-device (H2D) transfers are limited by $\sim$384GB/s in total, while individual D2H/H2D can each achieve higher bandwidth.

Despite these findings, the large disparity between achievable ($\sim$200GB/s) and achieved ($\sim$10GB/s) bandwidth raises a natural question: what prevents current LLM offloading mechanisms from fully harnessing GH200's NVLink-C2C?

\textbf{\uline{Insight \#3}}: Superchips unlock the bandwidth essential for responsive offloading, but current LLM serving stacks exploit $<$5\% of it. 
Fully exploiting its architecture requires co-designed scheduling and memory-management mechanism for high-performance, SLO-aware LLM inference.

\section{System Design}
\label{sec:design}

Based on insights from \sref{sec:obs}, we propose \name, a high-performance LLM serving system optimized for memory-intensive, latency-sensitive tasks on GH200.
This section details \name's overview (\sref{subsec:design-overall}) and its two key components: \scheduler (\sref{subsec:design-scheduler}) and \swapper (\sref{subsec:design-swapper}).
\subsection{Overall Architecture}
\label{subsec:design-overall}

\begin{figure}[!t]
    \centering
    \includegraphics[width=\columnwidth]{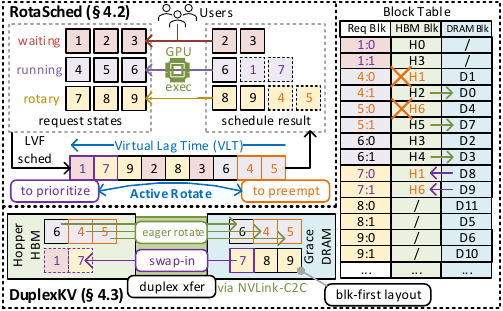}
    \caption{Overall architecture of \name. \scheduler maintains the requests states and perform Largest VLT First (LVF) scheduling based on Virtual Lag Time (VLT).
    \swapper manages the KV cache allocation across Hopper HBM and Grace DRAM with a block table, and transfers KV cache for swap-in and eager block rotation in a full-duplex manner. Numbers in squares denote different requests. ``X:Y'' denotes KV block Y of request X, ``H:W'' and ``D:Z'' denotes KV cache block in HBM and DRAM, respectively. }
    \label{fig:design-overall-arch}
\end{figure}

\fref{fig:design-overall-arch} illustrates the overall architecture of \name, which is composed of two co-designed components:

\noindent
\textbf{\scheduler} (\sref{subsec:design-scheduler}):
An OS-inspired, SLO-aware preemptive scheduler implementing \emph{active rotation} to fully utilize large swap space.
Guided by a novel \emph{Virtual Lag Time} (VLT) metric, it employs \emph{Largest-VLT-First} (LVF) policy to prioritize lagging requests vulnerable to SLO violations, while preempting long-running ones into the \emph{rotary} state for later resumption.
This ensures fast request execution rotation with offloading and high responsiveness for SLO violations.

\noindent
\textbf{\swapper} (\sref{subsec:design-swapper}):
A high-performance KV cache rotation engine optimized for Superchips, enabling \scheduler's frequent, large-volume offloading.
It achieves high bandwidth and low overhead via eager block rotation for data-race-free, full-duplex transfers, transformed KV cache layout with batched transfer kernel launches to merge small segments into efficient large transfers, and a cross-iteration pipeline to overlap transfers with computation.

\subsection{\scheduler: OS-Inspired Rotary Scheduler for SLO-Aware Offloading}
\label{subsec:design-scheduler}

\subsubsection{From Passive Preemption to Proactive Rotation}
\label{subsubsec:why-active-rotation}

Most existing LLM serving systems~\cite{kwon2023efficient,zheng2024sglang} employ \emph{passive preemption}: later-arrived requests are paused (by offloading/discarding their KV cache) only when KV cache demand exceeds available GPU memory.
While preventing out-of-memory (OOM), it fails to actively manage SLOs.
Crucially, no preemption occurs if GPU memory suffices for running requests, even if many waiting requests are close to violating their TTFT SLOs.

\begin{figure}[!t]
    \centering
    \includegraphics[width=\columnwidth]{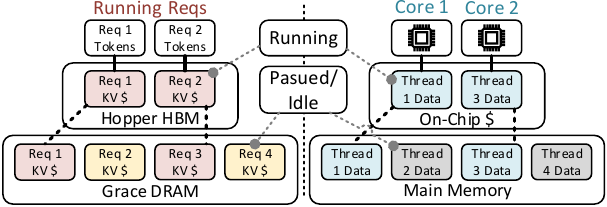}
    \caption{Analogy: LLM serving stack on GH200 (left) vs. OS on CPU (right). Requests $\to$ threads, Hopper HBM $\to$ on-chip cache, Grace DRAM $\to$ main memory, KV cache $\to$ thread data.}
    \label{fig:design-os}
\end{figure}

The GH200's tightly-coupled design offers a unique opportunity to rethink LLM latency and memory management.
As \fref{fig:design-os} illustrates, an LLM serving stack on GH200 is analogous to an OS on CPU: requests act as threads, Hopper HBM as on-chip cache, Grace DRAM as main memory, and KV cache as thread data.
This analogy inspires a shift in perspective: from static allocation of GPU memory to dynamic scheduling of GPU execution. 
Operating systems employ preemptive, time-slicing schedulers like CFS~\cite{wong2008towards} and EEVDF~\cite{stoica1995earliest} to rotate execution among numerous threads without starvation.

Based on this insight, \name adapts the \emph{rotation} principle for LLM serving to mitigate SLO violations.
Instead of treating preemption as a last resort to prevent OOM, \name \emph{actively rotates} requests across two execution tiers based on their SLO status.
It introduces a \emph{rotary} state: a transient execution state where a request's progress is temporarily paused on GPU and its KV cache is swapped to the CPU, waiting for next rotation.
This enables an LLM inference scheduling structure on Superchips analogous to OS time-slicing, but driven by SLO rather than fixed quanta, as introduced next. 
\fref{fig:design-overall-arch} shows request state transition in \name.

\subsubsection{Virtual Lag Time (VLT)}
\label{subsubsec:design-scheduler-vlt}

While the rotary state enables to alternate requests between GPU and CPU, the key question is \emph{when} rotation should occur.
Simply relying on queue length or memory usage would fall back to passive, SLO-unaware scheduling. 

A prominent lag-based thread scheduler in OS is \textit{Earliest Eligible Virtual Deadline First} (EEVDF)~\cite{stoica1995earliest}.
It tracks a \emph{lag} value that shows whether a thread has received its fair share of CPU time, and prioritizes those with the earliest virtual deadlines.
While offering valuable inspirations for managing a large number of concurrent requests, directly applying it to LLM serving is infeasible for two reasons. 
First, unlike lightweight context switches in OS supported by hardware-managed cache fetching, LLM request rotation involves costly GPU-CPU KV cache transfer.
Second, LLM serving must jointly manage two distinct, asymmetrically sensitive latency targets (TTFT and TBT).

\begin{figure}[!t]
    \centering
    \begin{minipage}[b]{0.48\columnwidth}
        \centering
        \includegraphics[width=0.9\linewidth]{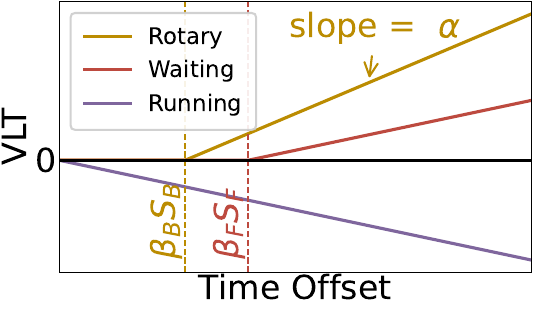}
        \caption{Visualization of VLT. The time offset refers to: $t_{\text{now}} - t_{\text{last}}$ (rotary), $t_{\text{now}} - t_{\text{arr}}$ (waiting), or  $t_{\text{now}} - t_{\text{run}}$ (running).}
        \label{fig:design-scheduler-vlt}
    \end{minipage}
    \hfill
    \begin{minipage}[b]{0.48\columnwidth}
        \centering
        \includegraphics[width=\columnwidth]{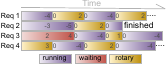}
        \caption{A conceptual example to show how LVF rotates executions. There are 4 requests in total, and HBM can only hold 2 requests. Numbers refer to VLT.}
        \label{fig:design-scheduler-rotate}
    \end{minipage}
\end{figure}

Building on EEVDF's spirit, \name introduces a metric called \emph{Virtual Lag Time} (VLT) to measure a request's deviation from its SLO progress.
VLT serves as the scheduling currency of \scheduler: requests with higher positive VLT are prioritized for execution, while those with smaller negative VLT become candidates for rotation. 
The following equation defines the VLT for each type of request:
\begin{equation*}
\resizebox{\columnwidth}{!}{$\displaystyle
    VLT = \begin{cases}
        \alpha \cdot \mathrm{ReLU} \left(t_{\text{now}} - t_{\text{last}} - \beta_B S_B \right), & \text{for rotary,} \\
        \mathrm{ReLU} \left(t_{\text{now}} - t_{\text{arr}} - \beta_F S_F \right), & \text{for waiting,} \\
        - \left(t_{\text{now}} - t_{\text{run}}\right), & \text{for running},
    \end{cases}
    $}
    \label{eq:vlt}
\end{equation*}
where $S_B$ and $S_F$ are the TTFT and TBT SLO, respectively. $\alpha\ge0$ is sensitivity weight, $\beta_B, \beta_F\in\mathbb{R}$ are latency tolerance coefficients for TTFT and TBT, respectively. $t_{\text{now}}$ is current system time, $t_{\text{last}}$ is the time of last generated token, $t_{\text{arr}}$ is request arrival time, $t_{\text{run}}$ is the time a request begins in the running state.
\fref{fig:design-scheduler-vlt} visualizes VLT for three request types when $\alpha>1$, $\beta_B>0$, $\beta_F>0$.
For running requests, VLT is negative and decreases as they run, with smaller values indicating higher degree of ``advance''.
Conversely, VLTs of waiting and rotary requests are initially zero when their inactive time is within tolerance.
They turn positive once tolerance is exceeded, increasing further with longer waits, reflecting their degree of ``lag''.

VLT provides a mechanism to monitor request time usage and potential SLO violations, enabling the SLO-aware priorities.
Intuitively, a larger VLT signifies more ``lag,'' thus requires higher execution priority to mitigate SLO violations.
Additionally, the adjustable parameters $\alpha$, $\beta_B$, $\beta_F$ offer flexibility for scenario-specific trade-offs.
First, $\alpha$ defines the sensitivity ratio of TBT/TTFT to SLO violations; larger values indicate higher relative TBT sensitivity.
Typically, TBT is more delay-sensitive (100ms is significant for TBT while manageable for TTFT),  suggesting $\alpha\ge1$.
Second, $\beta_B$ and $\beta_F$ reflect SLO requirements and tolerances in scheduling, where larger values indicate higher tolerance for SLO violations, enabling scenario-specific configuration.

\subsubsection{Largest-VLT-First (LVF) Scheduling}
\label{subsubsec:design-scheduler-flow}

\begin{figure}[!t]
    \centering
    \includegraphics[width=\columnwidth]{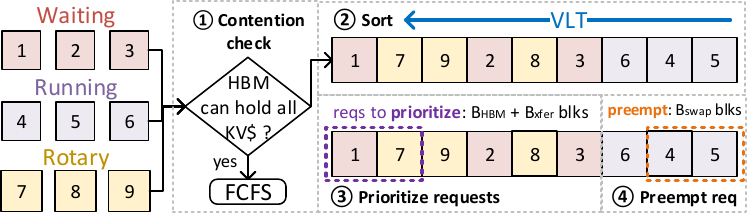}
    \caption{Demonstration of LVF scheduling algorithm. \ding{172}\ding{173}\ding{174}\ding{175} refer to steps described in \sref{subsubsec:design-scheduler-flow}.}
    \label{fig:design-scheduler-flow}
\end{figure}

Building on VLT, we next introduce the \textit{Largest-VLT-First} (LVF) scheduling policy, which leverages VLT to orchestrate proactive GPU-CPU request rotation.
LVF prioritizes requests by VLT:
those with higher positive VLT (indicating more vulnerable to SLO violations) are prioritized for execution, while those with lower negative VLT (indicating prolonged execution time) become preemption candidates to free HBM.
\fref{fig:design-scheduler-flow} illustrates the LVF algorithm (\aref{alg:ioaware}), which performs the following steps during each engine iteration:
\ding{\numexpr172} \textbf{Contention Check}: If HBM can hold KV cache of all requests, LVF skips subsequent steps and falls back to FCFS.
\ding{\numexpr173}
\textbf{Sort}: LVF calculates VLTs for all requests and sorts them in descending order as a list $\mathcal{L}$.
Waiting and rotary requests vulnerable to SLO violations are at the head.
The tail holds running requests which are preemption candidates.
\ding{\numexpr174}
\textbf{Prioritize requests}: Given a transfer budget $B_{\text{xfer}}$ and free HBM KV block number $B_{\text{HBM}}$, LVF picks rotary/waiting requests starting from the head of $\mathcal{L}$, until all $B_{\text{xfer}}+B_{\text{HBM}}$ blocks are used.
Selected requests are prioritized for execution.
\ding{\numexpr175}
\textbf{Preempt requests}: $B_{\text{swap}}$ extra blocks besides the current free HBM blocks are required to hold prioritized requests.
LVF starts from the tail of $\mathcal{L}$ to preempt running requests until getting enough extra blocks.

This algorithm allows LVF to dynamically monitor potential SLO violations using VLT, and perform preemptive scheduling.
By preempting (and swapping out) running requests, it makes HBM space for resuming SLO-vulnerable waiting or rotary requests.
Thus, LVF enables the SLO-aware use of large swap spaces, avoiding the TBT violations in naive policies (\sref{subsec:obs-why-need-scheduler}).
\fref{fig:design-scheduler-rotate} shows a conceptual example of how LVF rotates execution among 4 requests when HBM can only hold 2 requests.
Furthermore, the transfer budget $B_{\text{xfer}}$ controls the number of blocks to be swapped each iteration, reflecting the swap bandwidth.
NVLink-C2C enables a large $B_{\text{xfer}}$ and a high swap bandwidth compared to PCIe-based systems, which is critical to system responsiveness for backlogged waiting requests, as discussed in \sref{subsec:obs-pcie-vs-gh200}.

\begin{algorithm}[!t]
  \caption{LVF Scheduling}
  \label{alg:ioaware}
  \begin{algorithmic}[1]
    \Require Request (running $\mathcal{Q}_{\text{R}}$, waiting $\mathcal{Q}_{\text{W}}$, rotary $\mathcal{Q}_{\text{S}}$), KV cache block number of request $\mathrm{blk}(\cdot)$, transfer budget $B_{\text{xfer}}$, current free HBM block number $B_{\text{HBM}}$.
             
    \Ensure Preempted requests $\mathcal{P}$, prioritized requests $\mathcal{R}$.


    \State $\mathcal{P}\gets\emptyset$, $\mathcal{R}\gets\emptyset$, $B_{\text{left}} \gets B_{\text{HBM}} + B_{\text{xfer}}$
    \State $\mathcal{L} \gets \mathcal{Q}_{\text{R}} \cup \mathcal{Q}_{\text{W}} \cup \mathcal{Q}_{\text{S}}$ \Comment{All requests in a list}


    \If{$B_{\text{HBM}} \ge \sum_{r\in \mathcal{Q}_{\text{W}} \cup \mathcal{Q}_{\text{S}}} \mathrm{blk}(r)$} \Comment{\textbf{Step \ding{\numexpr172}}}
        \State \Return $\emptyset,\mathcal{Q}_{\text{W}} \cup \mathcal{Q}_{\text{S}}$ \Comment{Fallback to FCFS}
    \EndIf

    
    \State Sort $\mathcal{L}$ in descending order of $\mathrm{VLT}$. \Comment{\textbf{Step \ding{\numexpr173}}}


    \For{$r \in \mathcal{L}$} \Comment{Find prioritized requests} \Comment{\textbf{Step \ding{\numexpr174}}}
        \If{$\mathrm{VLT}(r) \ge 0 \And \mathrm{blk}(r) \le B_{\text{left}}$}
            \State $\mathcal{R} \gets \mathcal{R} \cup \left\{ r \right\}$, $B_{\text{left}} \gets B_{\text{left}} - \mathrm{blk}(r)$
        \EndIf
    \EndFor

    
    \State $B_{\text{swap}} \gets B_{\text{xfer}} - B_{\text{left}}$ \Comment{Extra block number} \Comment{\textbf{Step \ding{\numexpr175}}}

    \For{$r \in \mathrm{reversed} (\mathcal{L})$} \Comment{Find preempted request} 
        \If{$\mathrm{VLT}(r) < 0 \And B_{\text{swap}} >0$}
            \State $\mathcal{P} \gets \mathcal{P} \cup \left\{ r \right\}$, $B_{\text{swap}} \gets B_{\text{swap}} - \mathrm{blk}(r)$
        \EndIf
    \EndFor

    \State \Return $(\mathcal{P},\mathcal{R})$ \Comment{Preempted and prioritized requests}
  \end{algorithmic}
\end{algorithm}

\subsection{Efficient Offloading and Memory Management for Superchip NVLink-C2C}
\label{subsec:design-swapper}

\subsubsection{Why NVLink-C2C Utilization Remains Low?}
\label{subsubsec:why-under-utilization}

As discussed in \sref{subsec:obs-gh200-challenges}, existing offloading mechanisms suffer from severe under-utilization on NVLink-C2C, achieving only $\sim$10GB/s effective throughput, far below the theoretical peak.
To understand why, we examine how modern LLM serving frameworks manage KV cache.

\begin{figure}[!t]
    \centering
    \begin{minipage}[b]{0.48\columnwidth}
        \centering
        \includegraphics[width=\linewidth]{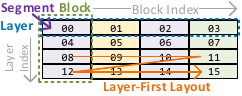}
        \caption{Layer, block, and segment structure of the PagedAttention KV cache. Colors denote different requests, and numbers denote relative segment addresses.}
        \label{fig:design-what-is-seg}
    \end{minipage}
    \hfill
    \begin{minipage}[b]{0.48\columnwidth}
        \centering
        \includegraphics[width=\linewidth]{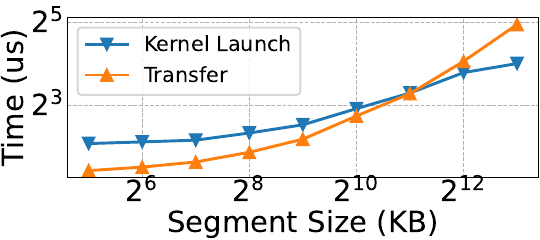}
        \caption{Kernel launch and transfer times for segments of different sizes via \texttt{cudaMemcpyAsync} (vLLM, Qwen2.5-32B, ShareGPT).}
        \label{fig:design-nvprof}
    \end{minipage}
\end{figure}

Most recent LLM serving systems employ PagedAttention~\cite{kwon2023efficient} to manage KV cache.
While eliminating memory fragmentation by allocating KV cache in fixed-sized blocks (e.g., 16 tokens per block), it also scatters each request's KV cache to non-contiguous GPU memory regions.
Supposing a Transformer-based model has $N_{\text{L}}$ layers, each maintaining their own KV cache.
During inference, each layer's KV cache is divided into blocks (with KV of $P$ tokens per block).
KV entries within the same layer and block form the largest contiguous memory region, which we call a segment (\fref{fig:design-what-is-seg}).
Segment size is $S_{\text{seg}} = P\times C$, where $C$ denotes per-token KV size.
Across $N_{\text{L}}$ layers and $N_{\text{B}}$ blocks, a request's full KV cache can be viewed as a layer-first 3D virtual tensor of shape $(N_{\text{L}}, N_{\text{B}}, S_{\text{seg}})$.

However, while KVs of the same segment are contiguous, segments from different layers or blocks are not, leading to non-contiguous memory layout. 
Since these segments are typically fine-grained and much smaller than transfer-efficient sizes, it results in poor C2C bandwidth utilization.
For example, in Qwen2.5-32B ($N_{\text{L}}=64$, $C=4$, $P=16$), each segment is $S_{\text{seg}} =64$KB, while a full block's KV cache (all 64 layers) is 4MB.
Thus, each 64KB segment must be transferred independently via a separate \texttt{cudaMemcpyAsync} kernel launch. 
As \fref{fig:bg-bw} shows, NVLink-C2C achieves high throughput ($\sim$200GB/s per direction) only when the transfer size $\ge$8MB.
Below that threshold, bandwidth drops sharply, falling $<$10GB/s for $\le$64KB segments.
Therefore, while PagedAttention efficiently utilizes GPU memory, it paradoxically keeps the NVLink-C2C operating in its low-efficiency regime.
Furthermore, the vast number of transfer operations also contributes to poor utilization.
Each \texttt{cudaMemcpyAsync} is launched internally as a lightweight GPU kernel.
When thousands of them are issued back-to-back, their accumulated launch overhead dominates execution time, further limiting effective bandwidth.
Our profile on GH200 using NVLink-C2C (\fref{fig:design-nvprof}) confirms that the current offloading path under-utilizes NVLink-C2C also due to the software induced granularity and launch overhead. 
Specifically, launch time of \texttt{cudaMemcpyAsync} is larger than transfer time when $S_{\text{seg}}\le4$MB.

\vspace{-0.5em}
\subsubsection{\swapper: Efficient KV Cache Rotation Engine}
\label{subsubsec:design-swapper-kernel}

Our analysis shows that NVLink-C2C bandwidth under-utilization stems not from hardware, but from thousands of small, non-contiguous \texttt{cudaMemcpyAsync} launches by current LLM serving stack.
To fully exploit the C2C, \name introduces \swapper, a high-performance KV cache rotation engine that (i) performs data-layout transformation to reorganize fragmented KV segments into large, contiguous regions optimized for transfer; (ii) pipelines computation with simultaneous D2H/H2D transfers, maximizing utilization of NVLink-C2C's full-duplex bandwidth.

\begin{figure}[!t]
    \centering
    \includegraphics[width=0.95\columnwidth]{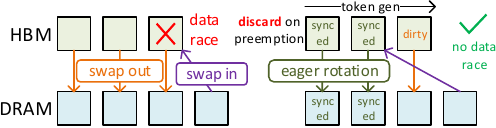}
    \caption{Left figure shows how data races occur when the swap-in destination block is the swap-out source block. Right figure shows our eager block rotation, which breaks the data dependency to enable concurrent swap-in and swap-out.}
    \label{fig:design-data-race}
\end{figure}

\noindent
\textbf{The challenge lies in swap-in/out dependencies.}
A straightforward bandwidth improvement is launching independent CUDA streams for concurrent H2D and D2H transfers.
However, naively overlapping them introduces data races on shared HBM blocks.
As \fref{fig:design-data-race} (left) shows, the destination HBM blocks for swap-in might coincide with the source blocks being freed by swap-out.
This dependency forces the swap-in stream to wait for swap-out completion, causing serialization of both directions and C2C link under-utilization.
While many prior systems discuss overlapping transfer with computation~\cite{gao2024cost,jiang2024neo}, few explicitly examine overlapping of bidirectional H2D/D2H transfers; for example, vLLM~\cite{kwon2023efficient} and SGLang~\cite{zheng2024sglang} perform swap-in and swap-out serially.
To our knowledge, no prior work has analyzed or addressed the data-race dependencies from bidirectional transfer concurrency in tightly coupled GPU-CPU systems like GH200.
This motivates our new \emph{eager block rotation} mechanism in \name, which eliminates these dependencies and enables full-duplex utilization of C2C.

\noindent
\textbf{Eager block rotation for data race free overlapping.}
To decouple two streams, \swapper introduces an \emph{eager block rotation} mechanism (\fref{fig:design-data-race} right).
The key insight is that KV cache generation is incremental: each request appends new tokens to its designated block sequentially, and previously fully written blocks remain unchanged until request completes.
Therefore, blocks are categorized: (i) \textit{Dirty}: partially filled and updated during token generation, and (ii) \textit{Synced}: fully filled and will not receive further writes until the request finishes.
\swapper eagerly swap-out \textit{synced} blocks from HBM and DRAM in the background, even before preemption.
These early transfers mark the corresponding HBM blocks as synced.
When the request is later preempted, only the last dirty block needs swap-out, while synced blocks in HBM can be discarded since valid copies already exist in DRAM.
\revised{
Crucially, this ``discarded on preemption'' mechanism relies on CPU-side duplication to maintain data correctness, ensuring that it imposes no additional HBM pressure.
}
A lightweight block table (as shown in \fref{fig:design-overall-arch}) tracks each block's state and residency.

\begin{figure}[!t]
    \centering
    \includegraphics[width=\columnwidth]{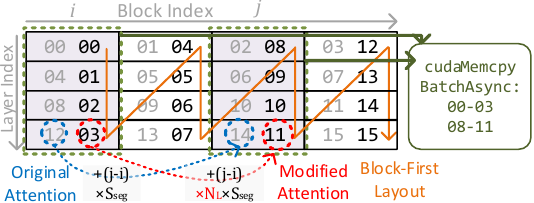}
    \caption{With block-first layout, small segments of the same block are merged into a larger contiguous region. Transfers of these regions can be done via a single \texttt{cudaMemcpyBatchAsync} kernel launch. Grey numbers are relative memory address for layer-first layout, black numbers are those for block-first layout.}
    \label{fig:design-merge}
\end{figure}

\noindent
\textbf{Block-first layout and batched KV transfers.}
To address the inefficiency of small, scattered memory segments, we redesign the KV cache storage from \emph{layer-first} to \emph{block-first} ordering.
As \fref{fig:design-merge} shows, this layout makes all layers within the same block contiguous, merging $N_{\text{L}}$ small per-layer segments into one large contiguous region (e.g., 64KB $\to$ 4MB for Qwen-2.5-32B).
This transformation converts numerous fine-grained transfers into much reduced large DMA operations, which are in the high-bandwidth regime of NVLink-C2C. 
We also extend PagedAttention to adopt this block-first layout.
As \fref{fig:design-merge} shows, for original PagedAttention with layer-first KV cache layout, the stride between block-$i$ and block-$j$ is $(j-i) \cdot S_{\text{seg}}$, while for block-first layout, it becomes $(j-i) \cdot N_L \cdot S_{\text{seg}}$.
We extend PagedAttention kernels to support this layout and stride, while preserving the paging abstraction. 
To further reduce kernel launch overhead, we merge individual \texttt{cudaMemcpyAsync} kernels of one direction (e.g., HBM $\to$ DRAM) into a single \texttt{cudaMemcpyBatchAsync} kernel.
This batched invocation issues all transfer descriptors in one kernel launch, effectively eliminating per-kernel launch overhead. 

\noindent
\textbf{Cross-iteration pipeline to overlap execution with rotation.}
To further reduce critical path latency, we also overlap schedule and KV cache rotation with model execution (e.g., LLM decoding), as \fref{fig:design-swapper-overlap} shows.
It employs a cross-iteration pipeline: during iteration-$t$, GPU executes the batch prepared in iteration-$(t-1)$, while the scheduler and \swapper concurrently prepare the batch for iteration-$(t+1)$ on the host side.
This pipeline hides data transfer and schedule latency behind compute, which minimizes GPU stall time and maximizing GPU utilization.

\section{Evaluation}
\label{sec:eval}

\subsection{Evaluation Methodology}
\label{subsec:eval-method}

\noindent
\textbf{Models and Workloads.}
We evaluate \name over three models: LLaMA-3-8B~\cite{dubey2024llama}, Qwen2.5-32B~\cite{yang2025qwen2} and Mixtral-8x7B~\cite{jiang2024mixtral}.
We use ShareGPT~\cite{ShareGPTTeam2023} and LMSYS-Chat-1M~\cite{zheng2023lmsys} datasets with controlled request arrival rates with sampled request arrival intervals following the Poisson distribution.

\noindent
\textbf{Metrics.}
We evaluate \name with the SLO attainment rate, defined as the percentage of requests that satisfy the SLO thresholds for TTFT and TBT.

\noindent
\textbf{Implementation.}
We implement \name in Python and C++ on top of vLLM~\cite{kwon2023efficient} (v0.6.6.post1), a widely used production-level LLM inference framework.

\revised{
\noindent
\textbf{Hardware Testbeds.}
All experiments are performed on an NVIDIA GH200 NVL2 system.
Each GH200 Superchip features 144GB of HBM and 480GB of DRAM.
We utilize \texttt{numactl} to configure NUMA memory affinity, ensuring that all allocated memory resides on the same Superchip.
}


\begin{figure}[!t]
    \centering
    \includegraphics[width=0.85\columnwidth]{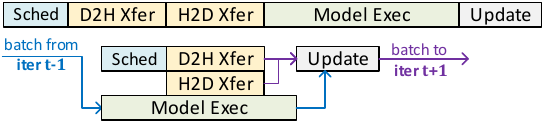}
    \caption{Comparing execution flow of vLLM (up) and \name (down). Schedule and KV cache transfers (2 CUDA streams) are overlapped with model execution in \name.}
    \label{fig:design-swapper-overlap}
\end{figure}
\begin{figure*}[!t]
    \centering
    \includegraphics[width=\linewidth]{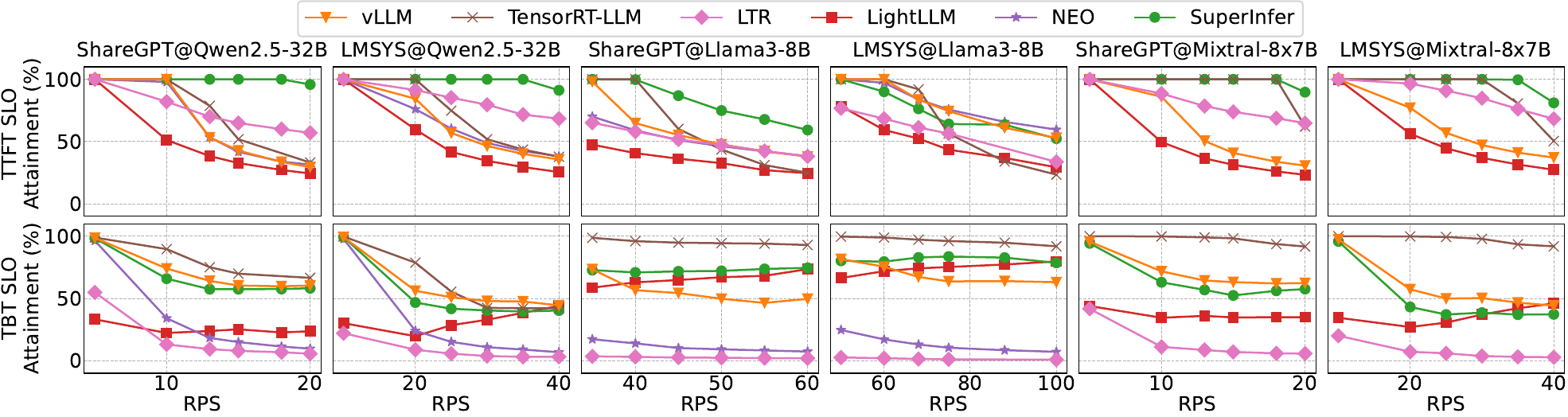}
    \caption{Comparison of \name against baselines across various models, datasets, and request rates (RPS). \name achieves significant improvements in TTFT SLO attainment over baselines, while preserving TBT SLO attainment comparable to others.}
    \label{fig:main-result}
\end{figure*}

\subsection{Main Result}
\label{subsec:eval-main}

We evaluate \name across all three models and all two datasets, comparing it against several baseline systems:

\noindent
\textbf{vLLM}~\cite{kwon2023efficient} (v0.6.6.post1), with V1 engine~\cite{vLLMV1vL65:online} enabled.
It is a widely-used production-level LLM inference system.

\revised{
\noindent
\textbf{TensorRT-LLM}~\cite{NVIDIATe66:online} (v1.1.0).
A highly optimized LLM inference framework by NVIDIA, incorporating a broad set of low-level and kernel optimizations.
}

\noindent
\textbf{LightLLM}~\cite{gong2025past}, \textbf{LTR}~\cite{fu2024efficient}: Two representative SLO-aware scheduler.
LightLLM balances trade-off between request queuing and harmful evictions by estimating future memory occupancy.
LTR approximates SJF using learning-based request length ranking.

\noindent
\textbf{NEO}~\cite{jiang2024neo}: A representative work to offload partial KV cache and Attention computation to the CPU.
Notably it does not support MoE models like Mixtral-8x7B.

For all baselines, PagedAttention~\cite{kwon2023efficient} and chunked prefill~\cite{agrawal2024taming} are enabled.
We exclude Pie~\cite{xu2024pie}, HeteGen~\cite{zhao2024hetegen}, and Select-N~\cite{ma2025memory} due to the lack of publicly available code.
FlexGen~\cite{sheng2023flexgen} is excluded for lacking essential features (PagedAttention, chunked prefill), which prevents fair comparison.
Approximate methods like InfiniGen~\cite{lee2024infinigen} are also excluded.

For our evaluation, we configure the \scheduler's parameters as: $\alpha = 3$, $\beta_B = 0$, $\beta_F = 0.5$, $B_{\text{xfer}}=2400$.
TTFT and TBT SLOs are set to 5s and 100ms, respectively.
Out of the 480 GB of Grace DRAM, we allocate 400 GB for KV cache offloading and conservatively reserve the remaining 80 GB for the OS and the \name runtime.
This 80 GB serves as a conservative safety margin, as \name introduces no extra memory consumption beyond the offloading space comparing to vLLM in our setup ($\sim$12 GB), leaving sufficient headroom for stable execution.
We report the TTFT and TBT SLO attainment rate under various request arrival rates.

\fref{fig:main-result} presents our results, revealing two key findings.
First, \name significantly surpasses baselines in TTFT SLO attainment, especially at high request rates.
This demonstrates its effectiveness of mitigating waiting queue HOL blocking via highly responsive active rotation, which proactively preempts running requests to prioritize waiting ones based on VLT (\sref{subsec:design-scheduler}).
Second, \name maintains TBT SLO attainment superior or comparable to baselines, indicating it leverages swap space in an SLO-aware manner for balanced TTFT and TBT improvement without introducing new rotary (swapped) queue HOL blocking (\sref{subsec:obs-why-need-scheduler}).
Among baselines, LTR achieves the best TTFT SLO attainment but significantly sacrifices TBT, as its static deadline-based priority mechanism fails to achieve the balanced improvements.
Notably, LightLLM shows unusual trends where TBT SLO attainment improves or stabilizes as the request rate increases, as it is designed to avoid harmful preemption caused by KV cache overflow and maintain stable TBT under high load (check Appendix \ref{app:lightllm} for further analysis).
\revised{
Furthermore, while TensorRT-LLM incorporates a broad set of advanced optimizations beyond KV-cache management and scheduling to achieve competitive TBT SLO attainment, its TTFT SLO attainment still degrades significantly at high request rates, due to its reliance on lazy preemption in request scheduling and its offloading strategy under memory pressure, which can delay prompt processing for newly arriving requests.
}

These results demonstrate \name's high performance and effectiveness in mitigating SLO violations.
This is achieved through the co-design of the SLO-aware, preemptive \scheduler and bandwidth-efficient \swapper.
Overall, \name exploits the full potential of the GH200 memory hierarchy and NVLink-C2C for SLO-aware offloading.

\subsection{Analysis Results}
\label{subsec:eval-analysis}

\subsubsection{How does each module of \name bring benefits?}

\begin{figure}[!t]
    \centering
    \includegraphics[width=\columnwidth]{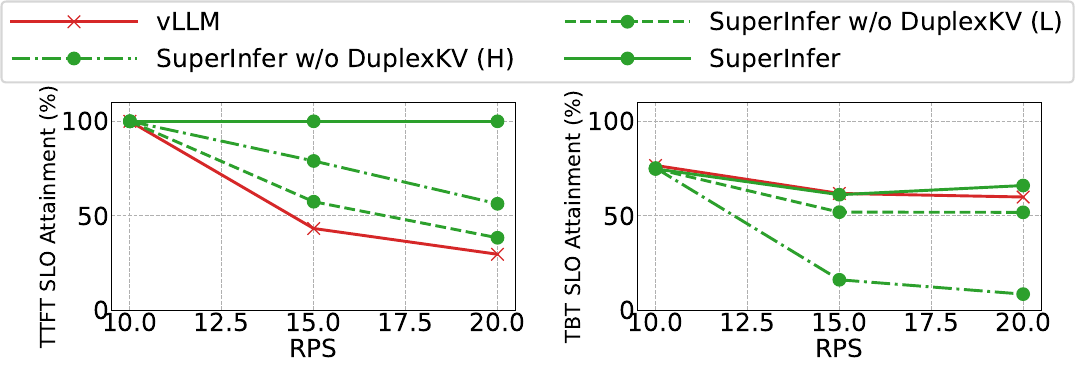}
    \caption{Comparing TTFT and TBT SLO attainment rates for: vLLM, \name w/o \swapper (L/H), \name.}
    \label{fig:eval-each-module}
\end{figure}

To isolate the benefit of each module proposed in \sref{sec:design}, we follow settings in \sref{subsec:eval-main} and evaluate the following configurations using the Qwen2.5-32B and ShareGPT dataset:
\textbf{(1) vLLM}: with default FCFS scheduler;
\textbf{(2) \name w/o \swapper}: \name with default KV cache offloading engine in vLLM, with two sub-variants for its transfer budget: \textbf{L} ($B_{\text{xfer}}=300$), which reflects the severe bandwidth under-utilization, and \textbf{H} ($B_{\text{xfer}}=2400$), which matches \swapper's budget;
\textbf{(3) \name}: The full version (\scheduler+\swapper).
\fref{fig:eval-each-module} shows the results.
Comparing vLLM and \name w/o \swapper (L) confirms that \scheduler effectively mitigates TTFT SLO violations while maintaining comparable TBT SLO attainment.
However, forcing a large $B_{\text{xfer}}$ with inefficient offloading engine as \name w/o \swapper (H) degrades performance, yielding significantly lower TBT SLO attainment.
This occurs because slow transfers become the critical bottleneck, preventing full overlap with model execution (\fref{fig:design-swapper-overlap}).
Finally, the full \name successfully manages the large $B_{\text{xfer}}$ using \swapper, achieving even greater TTFT improvements compared to \name w/o \swapper (L).

\subsubsection{How does \scheduler tame tradeoff between TTFT and TBT through VLT?}

\begin{figure}[!t]
    \centering
    \includegraphics[width=0.95\columnwidth]{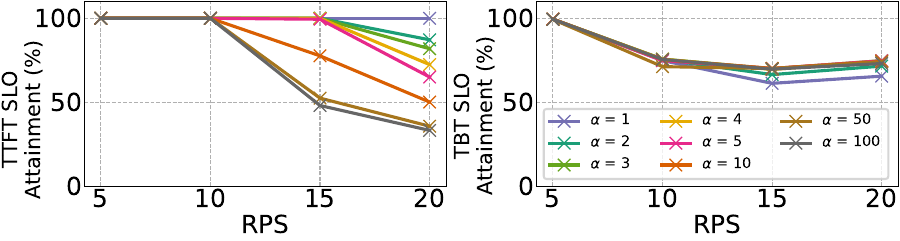}
    \caption{TTFT and TBT SLO attainment rate of \name under various $\alpha$. Larger $\alpha$ leads to better TBT but worse TTFT.}
    \label{fig:eval-vlt-slope}
\end{figure}

\begin{figure}[!t]
    \centering
    \includegraphics[width=0.95\columnwidth]{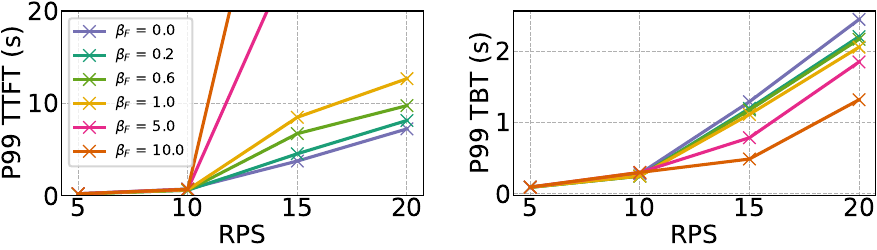}
    \caption{P99 TTFT and TBT of \name (various $\beta_F$ values).}
    \label{fig:eval-vlt-beta-ttft}
\end{figure}

\begin{figure}[!t]
    \centering
    \includegraphics[width=0.95\columnwidth]{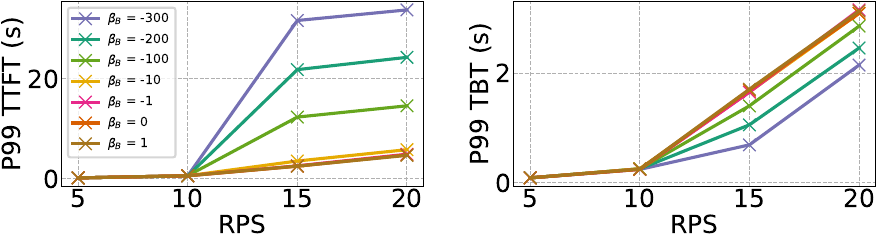}
    \caption{P99 TTFT and TBT of \name (various $\beta_B$ values).}
    \label{fig:eval-vlt-beta-tbt}
\end{figure}

As defined in \sref{subsubsec:design-scheduler-vlt}, VLT includes sensitivity parameter $\alpha$ for TTFT/TBT trade-off, and tolerance parameters $\beta_B$, $\beta_F$ to reflect SLOs and their tolerances.
We follow settings in \sref{subsec:eval-main} to evaluate their impacts using using Qwen2.5-32B and ShareGPT.
\fref{fig:eval-vlt-slope} shows results with varying $\alpha \ge1$ and fixed $\beta_B=\beta_F=0$.
It shows that larger $\alpha$ yields better TBT SLO attainment, as rotary requests get larger VLTs for prioritization.
However, this comes at the cost of lower TTFT attainment, as waiting requests are relatively delayed.
\revised{Thus, for TTFT-sensitive tasks (like summarization), we recommend setting $\alpha \le 1$ to get the best TTFT improvement.}
$\alpha=3$ offers a balanced sweet spot; further increases bring diminishing TBT benefits while significantly harming TTFT.
\fref{fig:eval-vlt-beta-ttft} shows fixing $\alpha=1$, $\beta_B=0$ and increasing $\beta_F$ leads to worse P99 TTFT, as new arrived requests have smaller VLTs and thus lower priority.
For $\beta_F\ge2$, P99 TTFT sharply increases with minimal TBT improvement, as new requests are significantly delayed.
Conversely, as \fref{fig:eval-vlt-beta-tbt} shows, when fixing $\alpha=1$, $\beta_F=0$ and varying $\beta_B$ in a large range, larger $\beta_B$ worsen P99 TBT, as rotary requests get smaller VLTs and lower priority for resumption.
Due to TBT's higher sensitivity to short delays, $\beta_B<0$ is suggested for best TBT tails latency, albeit at the cost of relatively higher TTFT tail latencies.
Overall, $\alpha$, $\beta_B$ and $\beta_F$ provide flexibility to trade off between TBT and TTFT improvements, enabling scenario-specific deployment tuning.

\revised{
It is important to note that while we provide a qualitative characterization to guide the understanding of parameter behavior under different settings, parameter selection can be tailored to specific applications.
\name does not attempt to correlate or predict query distributions when setting these parameters.
Such correlations are difficult to define in online LLM serving due to the dynamics and unpredictability of the workloads.
Instead, \name is designed to proactively react to KV cache pressure and SLO violation risks.
The exposed parameters ($\alpha$, $\beta_B$ and $\beta_F$) simply control how aggressively the system responds once such pressure is observed.
}


\subsubsection{How much bandwidth can \swapper achieve?}

To evaluate \swapper's optimizations for bandwidth utilization in \sref{subsec:design-swapper}, we measure the bandwidth and end-to-end (E2E) transfer time of bidirectionally transferring 16GB (8GB or 32768 tokens per direction) of Qwen2.5-32B KV cache.
We compare the following configurations:
(1) \textbf{Naive}: uses per-segment (64KB) copy as vLLM;
(2) \textbf{MS} (merged segments): uses our block-first layout to merge segments to 4MB, but still uses per-segment transfer;
(3) \textbf{MS+MK} (merged kernels): builds on MS, with merged transfer kernel per direction. Crucially, due to data races, transfers are still performed serially (one direction at a time);
(4) \textbf{\swapper}: the full version in \sref{subsubsec:design-swapper-kernel}, MS+MK+eager block rotation, enabling concurrent full-duplex transfers;
(5) \textbf{Ideal}: the theoretical limit of DRAM bandwidth (half-duplex 384GB/s, 192GB/s per direction concurrently).
\tref{tab:eval-bw} shows the results.
The naive method achieves only 5.6\% of ideal bandwidth and results in $37.4 \times$ the ideal E2E time, due to its high inefficiency with small-segment transfers (\sref{subsubsec:why-under-utilization}).
Instead, merging segments (MS) and kernels (MS+MK) incrementally improves realized uni-directional bandwidth within the limitation of DRAM (384GB/s), showing the importance of large transfer size and batch kernels.
However, due to data races, they can only perform direction-serialized transfers, thus achieving only $3.6\times$ and $1.4\times$ the ideal time, respectively.
In contrast, our proposed \swapper achieves near-ideal bidirectional bandwidth and E2E time consumption ($1.1\times$ the ideal), due to its eager block rotation to eliminate data race and enable concurrent bidirectional transfers.

\begin{table}[!t]
    \caption{Measured bandwidth and E2E time of different transfer engines. U denotes uni-directional, B denotes bidirectional.}
    \label{tab:eval-bw}
    \vspace{-10pt}
    \begin{center}
        \resizebox{\columnwidth}{!}{
            \begin{sc}
                \begin{tabular}{cccc}
                    \toprule
                    Method & D2H (GB/s) & H2D (GB/s) & E2E Time (ms) \\
                    \midrule
                    Naive  & 10.75(u) & 9.86(u) & 1556.15\\
                    MS & 80.05(u) & 133.51(u) & 159.87 \\
                    MS+MK & 238.95(u) & 269.69(u) & 63.14 \\
                    \textbf{\swapper} & \textbf{180.99(b)} & \textbf{179.37(b)} & \textbf{46.80} \\
                    Ideal (duplex)  & 192.00(b) & 192.00(b) & 41.66 \\
                    \bottomrule
                \end{tabular}
            \end{sc}
        }
    \end{center}
\end{table}

\subsubsection{How important is the high swap bandwidth?}

\begin{figure}[!t]
    \centering
    \includegraphics[width=0.95\columnwidth]{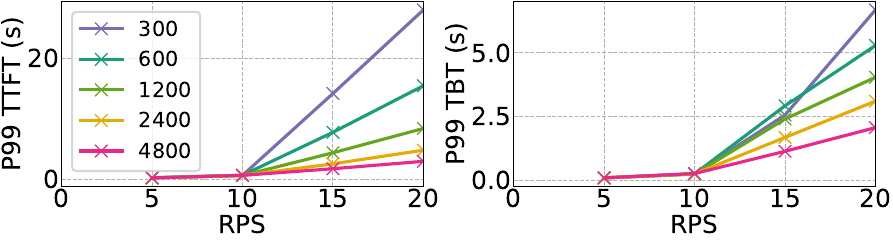}
    \caption{Comparing P99 TTFT and TBT of \name with various $B_{\text{xfer}}$. Higher $B_{\text{xfer}}$ significantly reduce tail latencies.}
    \label{fig:eval-swap-rate}
\end{figure}

In LVF scheduling (\sref{subsubsec:design-scheduler-flow}), the transfer budget ($B_{\text{xfer}}$) determines block number rotated per iteration, reflecting the swap bandwidth.
As discussed in \sref{subsec:obs-pcie-vs-gh200}, a high swap bandwidth is crucial to clear accumulated waiting requests timely and prevent new HOL blocking for the rotary (swapped) requests.
We validate this by testing \name with Qwen2.5-32B and ShareGPT, varying $B_{\text{xfer}}$ while keeping other settings same as \sref{subsec:eval-main}.
The results in \fref{fig:eval-swap-rate} show that a larger $B_{\text{xfer}}$ significantly reduces the P99 TTFT and TBT, which mainly comes from the time requests spend in waiting or rotary states.
Also, larger budgets yield greater improvements, confirming the necessity of high swap bandwidth for effective SLO-aware LLM serving with offloading.

\revised{
A potential concern with \name is whether KV cache transfers might introduce pipeline stalls if they do not complete before model inference finishes.
Upon reviewing the data collected across all experiments in \sref{subsec:eval-main}, we find that incomplete overlaps which lead to execution stalls occurred in only 0.021\% of all engine iterations.
A detailed breakdown reveals average latencies of 7.63 ms for scheduling, 15.8 ms for KV transfers, and 69.82 ms for model execution.
Because the scheduling and transfer latencies are substantially shorter than the model execution time, their overheads are effectively hidden.
}

\begin{figure}[!t]
    \centering
    \includegraphics[width=0.95\columnwidth]{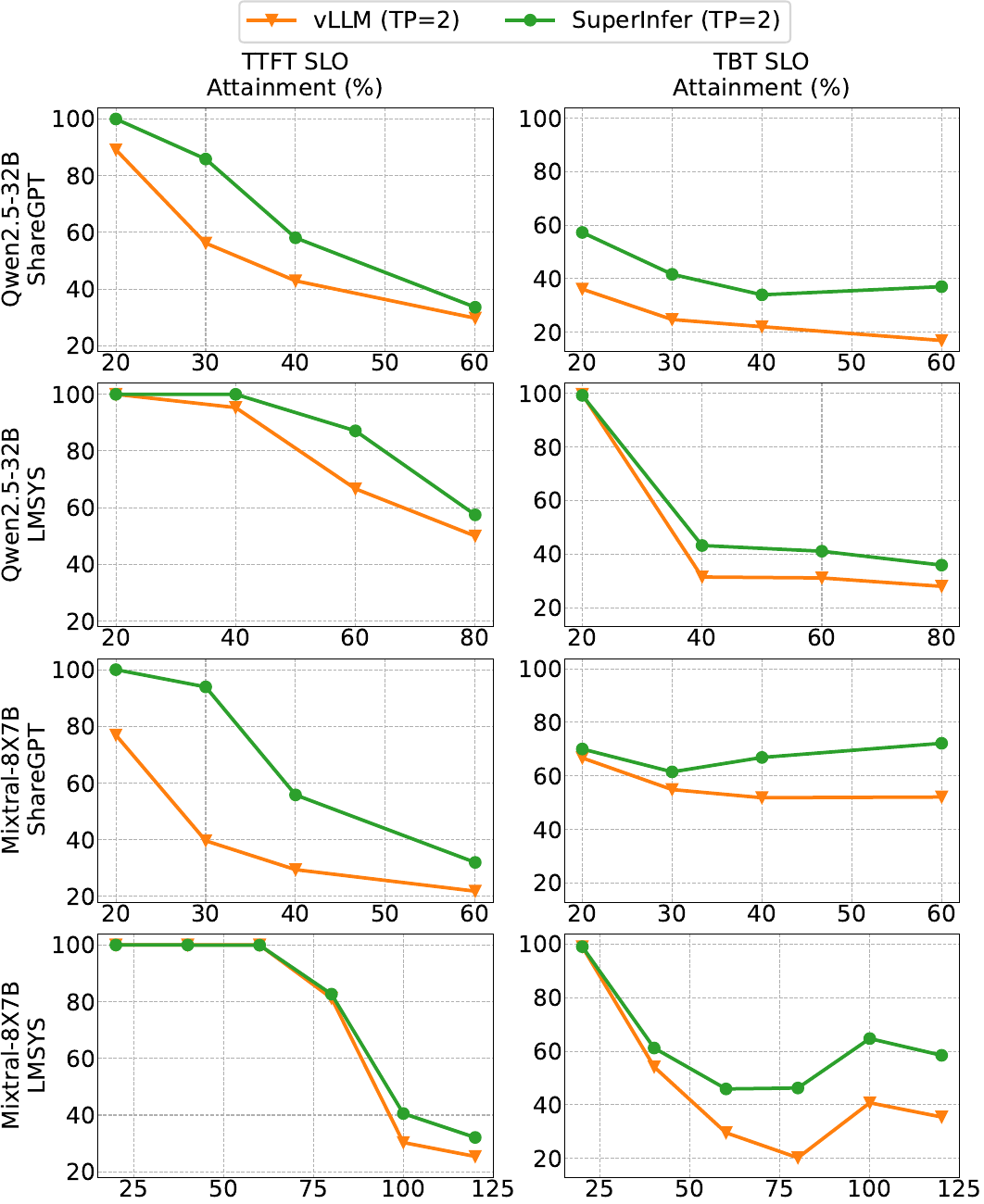}
    \caption{Comparison of \name and vLLM under TP=2 settings. The performance benefits of \name successfully extend to distributed configurations.}
    \label{fig:eval-tp}
\end{figure}

\revised{
\subsubsection{Does \name extend to multi-GPU settings?}

To confirm that \name's benefits persist in distributed environments, we evaluated it using Tensor Parallelism (TP=2) with Qwen2.5-32B and Mixtral-8x7B.
GPUs are connected with NVLink (900GB/s).
As \fref{fig:eval-tp} shows, across both models and datasets, \name consistently achieves higher TTFT and TBT SLO attainment rates compared to vLLM.
This is because our \scheduler and \swapper are fundamentally orthogonal to standard parallelism strategies; they introduce no extra inter-GPU traffic, allowing \name to scale effectively in multi-GPU and distributed setups.

}

\subsubsection{How does \name affect the throughput?}

\begin{figure}[!t]
    \centering
    \includegraphics[width=\columnwidth]{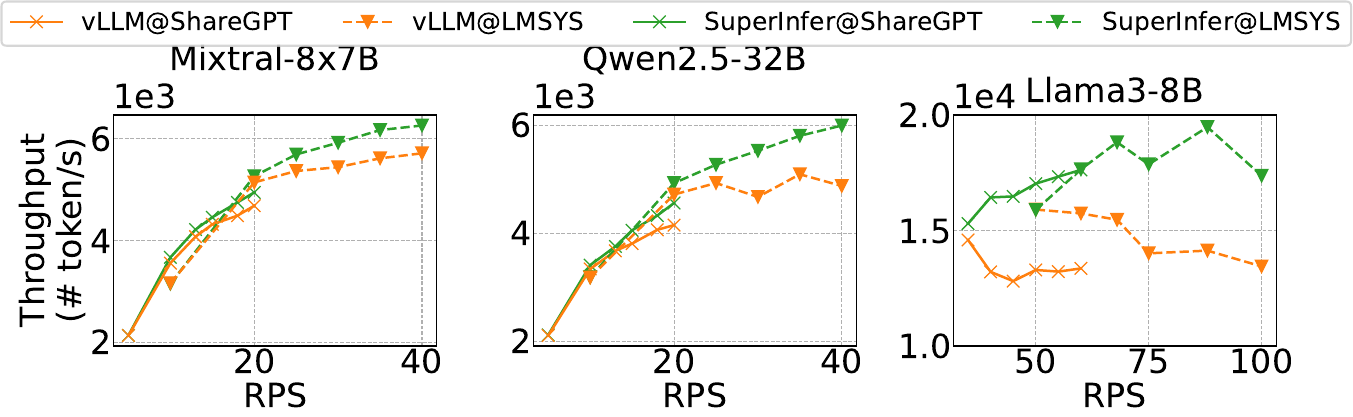}
    \caption{Throughput of vLLM and \name on three models.}
    \label{fig:eval-throughtput-goodput}
\end{figure}

\fref{fig:eval-throughtput-goodput} shows measured throughput of vLLM and \name from \sref{subsec:eval-main}.
\name achieves comparable or slightly better throughput than vLLM, with up to 29.2\% improvement at high request rates.
This is because in conventional offloading, memory-heavy long requests can monopolize GPU memory for a long time, preventing new requests from leveraging the concurrency benefit of chunked prefill~\cite{agrawal2024taming}, as prefill requests have few batching opportunities.
In contrast, \name's fast rotation offers prefill requests more batching opportunities.

\section{Conclusion}
\label{sec:conclusion}

This paper presents \name, a high-performance, SLO-aware LLM serving system optimized for Superchips.
\name transforms passive preemption into proactive, fine-grained request scheduling, via an OS-inspired active rotary scheduler and a co-designed \swapper rotation engine, achieving high NVLink-C2C utilization.
Evaluations across diverse models and workloads show that \name substantially improves SLO attainment under high request rates while maintaining throughput.
Our study demonstrates that Superchips unblock new opportunities for LLM serving, but realizing their full potential requires careful co-design of scheduling and memory movement in the software system.

\section*{Acknowledgments}

We sincerely appreciate the anonymous reviewers and our shepherd. Their insightful feedback helps significantly improve the quality of the paper. This research was supported by the National Science Foundation (NSF) under Grant No. 2441601. The work utilized the Delta and DeltaAI system at the National Center for Supercomputing Applications (NCSA) and Jetstream2 at Indiana University through allocation CIS240055 from the Advanced Cyberinfrastructure Coordination Ecosystem: Services \& Support (ACCESS) program, which is supported by National Science Foundation grants \#2138259, \#2138286, \#2138307, \#2137603, and \#2138296. The Delta advanced computing resource is a collaborative effort between the University of Illinois Urbana-Champaign and NCSA, supported by the NSF (award OAC 2005572) and the State of Illinois. UIUC SSAIL Lab is supported by research funding and gift from Google, IBM, Amazon, and AMD, including the Google ML and Systems Junior Faculty Award.

\clearpage

\bibliography{_s99_ref}
\bibliographystyle{mlsys2025}

\clearpage
\appendix
\section{Comparing FCFS and SJF-Oracle}
\label{app:sjf}

As discussed in \sref{subsec:obs-why-need-scheduler}, recent LLM serving systems increasingly employ sophisticated scheduling techniques to manage latency.
However, they cannot overcome the memory constraints posed by hardware.
Once KV cache storage is exhausted, newly arrived requests have to backlog in the waiting queue, waiting for running ones to finish.

\begin{figure}[!ht]
    \centering
    \includegraphics[width=0.7\linewidth]{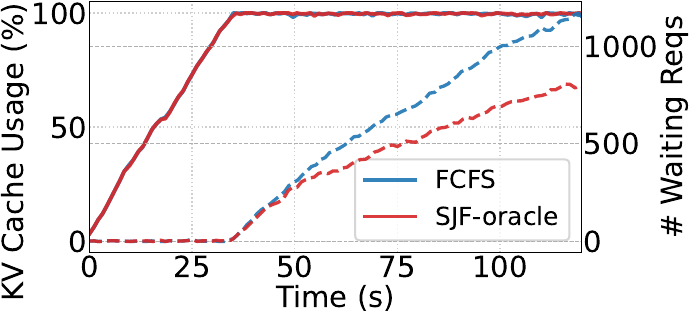}
    \caption{KV cache usage and waiting request number for vLLM with FCFS and SJF-oracle scheduler. Model: Qwen2.5-32B, dataset: ShareGPT, RPS=20.}
    \label{fig:bg-starvation-example}
\end{figure}

To illustrate, we measure vLLM on GH200 using Qwen2.5-32B~\cite{yang2025qwen2} model and ShareGPT dataset~\cite{ShareGPTTeam2023}.
We compare the \emph{First-Come-First-Serve} (FCFS) and \emph{Shortest-Job-First} with oracle generation length information (SJF-Oracle) policy.
As shown in \fref{fig:bg-starvation-example}, both FCFS and SJF-Oracle fail to prevent TTFT SLO violations under memory pressure.
Once KV cache storage is exhausted, the length of waiting queue spikes dramatically, which causes severe TTFT SLO violations.

\section{Bandwidth Measurement}
\label{app:bw-measure}

We measure the CPU-GPU bandwidth for GH200 and H200 using NVIDIA's open-source \href{https://github.com/NVIDIA/nvbandwidth}{\texttt{nvbandwidth}} tool (v0.8).
We focus on two specific bidirectional copy engine (CE) test cases:
\begin{itemize}
    \item \texttt{host\_to\_device\_bidirectional\_memcpy\_ce} (Test ID 2)
    \item \texttt{device\_to\_host\_bidirectional\_memcpy\_ce} (Test ID 3)
\end{itemize}

The following command is used to run all tests:

\begin{lstlisting}[language=bash]
CUDA_VISIBLE_DEVICES=0 numactl --cpunodebind=0 --membind=0 ./nvbandwidth -t 2 3 -b <size_in_MB> -i 3
\end{lstlisting}

\section{Further Analysis of LightLLM}
\label{app:lightllm}

As discussed in \sref{subsec:eval-main}, LightLLM~\cite{gong2025past} shows an unusual TBT SLO attainment rate trend: as request rates increase, the TBT SLO attainment rate improves or stabilizes.
To further explore the underlying root cause, we analyze the cumulative distribution function (CDF) of TBT for LightLLM.

\begin{figure}[!ht]
    \centering    
    \includegraphics[width=0.7\columnwidth]{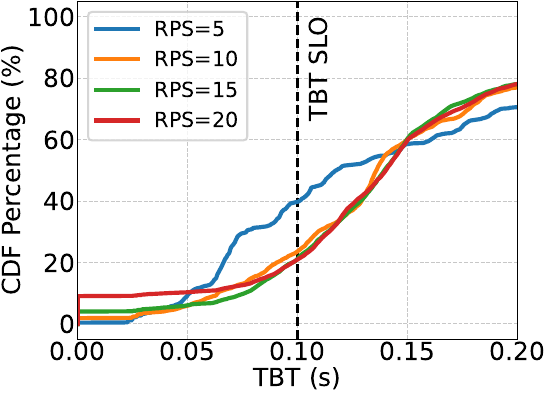}
    \caption{Cumulative distribution function (CDF) of TBT for LightLLM in \sref{subsec:eval-main}. When request rate increase, LightLLM }
    \label{fig:bg-naive-swapper}
\end{figure}

As shown in \fref{fig:bg-naive-swapper}, when request rate increase, the CDF of TBT of LightLLM shows almost no changes.
Therefore, the TBT SLO attainment rate also stabilizes.
That is because LightLLM's ``Past-Future'' scheduler is designed to avoid harmful request evictions by precisely estimating the peak future KV cache required by the running batch.
Unlike aggressive schedulers that underestimate memory consumption and suffer from frequent, harmful evictions under high load, LightLLM's approach ensures that scheduled requests can complete successfully without the response interruptions that typically cause TBT SLO violations.
This stable TBT performance, free from eviction-induced stalls, directly translates to a stabilized SLO attainment rate even as request rates increase.

\section{vLLM on Unified Memory of GH200}
\label{app:vllm-um}

\begin{figure}[!t]
        \centering    
        \includegraphics[width=0.6\columnwidth]{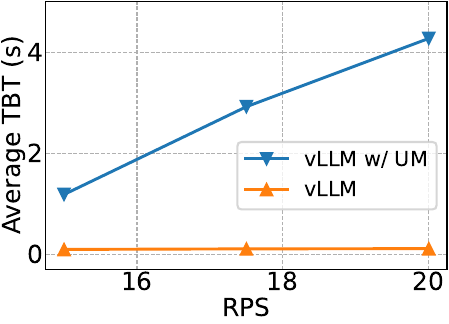}
        \caption{Comparing the vLLM and that with KV cache storage in GH200's Unified Memory (UM). vLLM on UM shows significant TBT degradation.}
        \label{fig:bg-um}
\end{figure}

Some readers may consider the hardware-managed Unified Memory (UM) of GH200~\cite{gh200} as a potential good solution for offloading.
UM merges HBM and DRAM into a unified address space and automatically migrates blocks based on access, allowing for a unified large KV cache storage without explicit offloading management.
However, our test result in \fref{fig:bg-um} show that enabling UM for vLLM severely degrades performance, leading to significantly higher average TBT.

This performance collapse stems not from traditional UM overheads caused by page-fault-driven migration, but from a fundamental architecture and workload mismatch.

On conventional PCIe-based GPUs, UM is page-fault driven.
An access to a non-resident page (e.g., in CPU memory) triggers a costly page fault and GPU stall, after which the driver migrates the page over PCIe.
In contrast, the GH200's architecture features a cache-coherent NVLink-C2C interconnect and Address Translation Services (ATS)~\cite{gh200}.
This allows the Hopper GPU to directly access the Grace CPU's DRAM without incurring any page faults.

GH200 does support page migration, but instead of being page-fault driven, it uses \emph{hardware access counters} to track the access frequency of pages from both the CPU and GPU.
When a CPU-resident page is accessed from the GPU side, it initially keeps its residency in CPU memory.
Only after enough frequent accesses will the GH200 trigger a background migration from CPU to GPU.

Although this migration is transparent to applications like vLLM and involves no page faults, it exposes a new bottleneck: the bandwidth cliff.
The GPU's local HBM memory provides bandwidth up to 4 TB/s, whereas the CPU's DRAM subsystem, even when accessed directly over the 900 GB/s C2C link, is limited to its own lower bandwidth of 384 GB/s.
This makes GPU access to CPU memory nearly an order of magnitude slower than access to HBM.

This bandwidth cliff is the root cause of vLLM's poor performance on UM.
When a new request's KV cache blocks are resident in DRAM, the GPU's initial accesses are all serviced over the low-bandwidth C2C link through ATS, leading to poor performance in the Attention kernel.
The request may finish processing long before the hardware counters accumulate enough frequency to ever trigger a migration.
The system is thus perpetually ``warming up'' data that is already ``cold'' or no longer in active use.

Therefore, this hardware-managed UM is not suitable for LLM serving with offloading.

\clearpage
\section{Artifact Appendix}





\subsection{Artifact check-list (meta-information)}

{\small
\begin{itemize}
    \item {\bf Algorithm: } \scheduler: Largest-VLT-First (LVF) Scheduling; \swapper: Eager Block Rotation.
    \item {\bf Program: } \name (implemented in Python and C++, atop vLLM v0.6.6.post1\footnote{\label{vllm}\url{https://github.com/vllm-project/vllm/releases/tag/v0.6.6.post1}}).
    \item {\bf Compilation: } CUDA 12.8 and GCC 13.3.0 for \swapper and CUDA kernels of \name.
    \item {\bf Dataset: } ShareGPT\footnote{\label{sharegpt}\url{https://huggingface.co/datasets/anon8231489123/ShareGPT_Vicuna_unfiltered/blob/main/ShareGPT_V3_unfiltered_cleaned_split.json}}; LMSYS-Chat-1M\footnote{\label{lmsys}\url{https://huggingface.co/datasets/lmsys/lmsys-chat-1m}}.
    \item {\bf Run-time environment: } Ubuntu 24.04.3 LTS with kernel 6.8.0-100-generic-64k, CUDA 12.8.
    \item {\bf Hardware: } NVIDIA GH200 NVL2\footnote{\label{gh200}\url{https://www.nvidia.com/en-us/data-center/grace-hopper-superchip/}}, with 2 Grace-Hopper pairs, each including 144GB HBM and 480GB DRAM connected via NVLink-C2C.
    \item {\bf Execution: } A pre-configured Docker environment with automated bash scripts for orchestrating server and client processes, benchmarking baselines, parsing execution logs, and generating performance plots..
    \item {\bf Metrics: } Time-To-First-Token (TTFT) SLO attainment rate; Time-Between-Tokens (TBT) SLO attainment rate; Throughput (tokens/s); Achieved NVLink-C2C transfer bandwidth (GB/s) and transfer time (ms).
    \item {\bf Output: } Parsed logs and Python-generated plots.
    \item {\bf Experiments: } End-to-end SLO attainment comparison against baselines (vLLM, LightLLM, LTR, NEO).
    \item {\bf Disk Requirement: } $\sim$500 GB (to accommodate model weights for Qwen2.5-32B, Mixtral-8x7B, LLaMA-3-8B, dataset caches, and the Docker image).
    \item {\bf Time Requirement: } Time for pulling docker image varies based on the network connection speed.
    The whole experiments takes about 30 hours.
    Because the full evaluation across all models and datasets is time-consuming, we also provide a ``lite'' experiment for convenience.
    This lite version evaluates a single model on a single dataset to quickly reproduce the core main results, which can be completed in approximately 5 hour.
    \item {\bf Code licenses: } Apache 2.0\footnote{\label{apache}\url{https://choosealicense.com/licenses/apache-2.0/}}.
    \item {\bf Data licenses: } LMSYS-Chat-1M Dataset License Agreement\footnote{\label{lmsts-license}\url{https://huggingface.co/datasets/lmsys/lmsys-chat-1m\#lmsys-chat-1m-dataset-license-agreement}} (LMSYS-Chat-1M) and Apache 2.0\footref{apache} (ShareGPT).
    \item {\bf Archived DOI: } Artifact: \path{https://doi.org/10.5281/zenodo.18971768}. Code: \path{https://doi.org/10.5281/zenodo.19394229}.
\end{itemize}

\subsection{Description}




\subsubsection{Hardware dependencies}
\label{subsubsec:hardware}

\begin{itemize}
    \item \textbf{Platform:} NVIDIA GH200 NVL2. This system features two Grace CPU and Hopper GPU pairs. Each Hopper GPU is equipped with 144GB HBM, and each Grace CPU contains 480GB DRAM (with a maximum bandwidth of 384GB/s). Within each pair, the Grace CPU and Hopper GPU are tightly coupled via an NVLink-C2C interconnect, providing 900GB/s of bidirectional bandwidth. Additionally, the two Hopper GPUs are interconnected via a 900GB/s NVLink.
    \item \textbf{Storage:} About 500GB of free disk space is required to accommodate the Docker image, model weights, and dataset cache.
\end{itemize}

\subsubsection{Software dependencies}

\begin{itemize}
    \item \textbf{OS \& Environment:} Ubuntu 24.04.3 LTS with kernel 6.8.0-100-generic-64k, CUDA 12.8 and GCC 13.3.0. Our provided Docker image is pre-packaged with this environment. The virtual environments for \name and all evaluated baselines are managed and isolated using Anaconda.
    \item \textbf{\name:} Implemented atop vLLM v0.6.6.post1\footnote{\label{vllm}\url{https://github.com/vllm-project/vllm/releases/tag/v0.6.6.post1}}, with PyTorch 2.5.1.
    \item \textbf{Baselines:} The evaluated baselines have the following specific software dependencies:
    \begin{itemize}
        \item \textbf{vLLM:} Version v0.6.6.post1\footref{vllm} with PyTorch 2.5.1.
        \item \textbf{LTR\footnote{\url{https://github.com/hao-ai-lab/vllm-ltr}}:} Git commit \texttt{\seqsplit{13bbf6ff3dab661791d41362551b089e5f77c91c}} with PyTorch v2.4.1.
        \item \textbf{LightLLM\footnote{\url{https://github.com/ModelTC/LightLLM}}:} Version v1.1.0 with PyTorch 2.9.0.
        \item \textbf{NEO\footnote{\url{https://github.com/NEO-MLSys25/NEO}}:} Git commit \texttt{\seqsplit{33e4a0f7632688e4122de4c3c140196cebea6a5a}} with PyTorch 2.7.0 and Triton 3.3.0\footnote{\url{https://github.com/triton-lang/triton}}.
    \end{itemize}
\end{itemize}

\subsubsection{Datasets}

We use two open-source datasets to evaluate \name and the baselines:

\begin{enumerate}
    \item \textbf{ShareGPT\footnote{\label{sharegpt}\url{https://huggingface.co/datasets/anon8231489123/ShareGPT_Vicuna_unfiltered/blob/main/ShareGPT_V3_unfiltered_cleaned_split.json}}:} This dataset contains highly diverse, real-world conversational turns between users and LLMs. In our evaluation, following prior works such as LTR, we extract only the first conversational turn (the initial human prompt and the LLM response) to serve as a single request.
    \item \textbf{LMSYS-Chat-1M\footnote{\label{lmsys}\url{https://huggingface.co/datasets/lmsys/lmsys-chat-1m}}:} A large-scale, real-world dataset comprising one million conversations collected in the wild from the LMSYS Chatbot Arena.
\end{enumerate}

During testing, for each dataset, we randomly sample a total of $120 \times \mathrm{RPS}$ requests (representing 120 seconds of traffic at a target request rate).
These requests are then dispatched to the server with inter-request delays sampled from a Poisson distribution to simulate realistic arrival patterns.

\subsubsection{Models}

We use three open-source LLMs for our evaluations to cover diverse model scales and architectural designs:

\begin{itemize}
    \item \textbf{LLaMA-3-8B\footnote{\url{https://huggingface.co/meta-llama/Meta-Llama-3-8B}}:} Represents widely adopted, small-scale dense models.
    \item \textbf{Mixtral-8x7B\footnote{\url{https://huggingface.co/mistralai/Mixtral-8x7B-v0.1}}:} Represents Mixture-of-Experts (MoE) architectures.
    \item \textbf{Qwen2.5-32B\footnote{\url{https://huggingface.co/Qwen/Qwen2.5-32B}}:} Represents larger-scale dense models.
\end{itemize}

\subsection{Installation}

To simplify the environment setup and installation process, we provide a self-contained Docker image.
This image comes pre-packaged with all necessary runtime dependencies, the \name itself, the evaluated baselines.
Model weights will be download automatically when running the scripts (Hugging Face token required).

\begin{lstlisting}[style=elegantbash]
docker pull monsoon235/superinfer_ae_public
docker run --gpus all -it -u 1003 -w /home/mingtao4 monsoon235/superinfer_ae_public zsh  # enter the environment
\end{lstlisting}



\subsection{Experiment workflow and expected result}

A full evaluation of the main results takes roughly 30 hours due to the extensive combination of 3 models, 2 datasets, 5 methods, and various request rates.
For a quicker reproduction, you can run the provided ``lite'' experiment ($\sim$5 hour), which evaluate all methods on just 1 model (Qwen2.5-32B) and 1 dataset (ShareGPT).

\subsubsection{Lite Version}

To execute the ``lite'' version of the experiment, please run the following commands:

\begin{lstlisting}[style=elegantbash]
ulimit -n 100000
conda activate vllm-v1
python /home/mingtao4/AE_scripts/AELite_AIO_0309_v1.py
\end{lstlisting}

This automated script evaluates \name alongside the baselines (vLLM, LTR, LightLLM, and NEO) using the Qwen2.5-32B model and the ShareGPT dataset.
It sequentially launches the LLM server and the benchmark client for all data points.
The generated logs are saved in \path{AE_results}.

To reproduce the figure in the paper, run Jupyter Notebook \path{/home/mingtao4/AE_scripts/AELite_mainresult.ipynb} to automatically parse these logs and reproduce the figure.

\textbf{Expected Results}: It should reproduce the first column (ShareGPT @ Qwen2.5-32B) of \fref{fig:main-result} in the paper.

\subsubsection{Full Version}

To execute the full version of the experiment, please run the following commands:

\begin{lstlisting}[style=elegantbash]
ulimit -n 100000
conda activate vllm-v1
python /home/mingtao4/AE_scripts/AE_AIO_0309_v1.py
\end{lstlisting}

This automated script evaluates \name alongside the baselines (vLLM, LTR, LightLLM, and NEO) using all models and datasets.
It sequentially launches the LLM server and the benchmark client for all data points.
The generated logs are saved in \path{AE_results}.

To reproduce the figure in the paper, run Jupyter Notebook \path{/home/mingtao4/AE_scripts/AE_mainresult.ipynb} to automatically parse these logs and reproduce the figure.

\textbf{Expected Results}: It should reproduce the full \fref{fig:main-result} in the paper.



\end{document}